\documentclass[amssymb, preprintnumbers, showpacs, showkeys, aps, prl, superscriptaddress, twocolumn, longbibliography]{revtex4-2}

\usepackage[hidelinks, hypertexnames=false]{hyperref}
\usepackage[all]{hypcap}
\usepackage{slashed}
\usepackage{graphicx}
\usepackage{amsmath}
\usepackage{latexsym}
\usepackage{epstopdf}
\usepackage{amsmath}
\usepackage{enumitem}
\usepackage{amssymb,amsmath}
\usepackage{multirow}
\usepackage{mathtools}
\usepackage{bbm}
\usepackage{bm}
\usepackage{amsfonts}
\usepackage{amssymb}
\usepackage{calligra}
\usepackage{calrsfs}
\usepackage{makecell}
\usepackage[normalem]{ulem}
\usepackage[USenglish,american]{babel}

\usepackage{xcolor}

\usepackage{physics}
\newcommand{\ave}[1]{\expval**{#1}}

\renewcommand{\ol}[1]{\overline{#1}}

\begin{document}

\title{Quasicrystalline Bose glass in the absence of disorder and quasidisorder}

\author{Matteo Ciardi}
\email{matteo.ciardi@unifi.it}
\affiliation{Dipartimento di Fisica e Astronomia, Universit\`a di Firenze, I-50019, Sesto Fiorentino (FI), Italy}
\affiliation{INFN, Sezione di Firenze, I-50019, Sesto Fiorentino (FI), Italy}

\author{Adriano Angelone}
\email{adriano.angelone@exact-lab.it}
\affiliation{Sorbonne Universit\'e, CNRS, Laboratoire de
Physique Th\'eorique de la Mati\`ere Condens\'ee, LPTMC,
F-75005 Paris, France}
\affiliation{eXact lab s.r.l., Via Francesco Crispi 56 - 34126 Trieste, Italy}

\author{Fabio Mezzacapo}
\email{fabio.mezzacapo@ens-lyon.fr}
\affiliation{Univ Lyon, Ens de Lyon, CNRS, Laboratoire de Physique, F-69342 Lyon, France}

\author{Fabio Cinti}
\email{fabio.cinti@unifi.it}
\affiliation{Dipartimento di Fisica e Astronomia, Universit\`a di Firenze, I-50019, Sesto Fiorentino (FI), Italy}
\affiliation{INFN, Sezione di Firenze, I-50019, Sesto Fiorentino (FI), Italy}
\affiliation{Department of Physics, University of Johannesburg, P.O. Box 524, Auckland Park 2006, South Africa}

\begin{abstract}
We study the low-temperature phases of interacting bosons on a two-dimensional quasicrystalline lattice. By means of numerically exact Path Integral Monte Carlo simulations, we show that for sufficiently weak interactions the system is a homogeneous Bose-Einstein condensate, which develops density modulations for increasing filling factor. The simultaneous occurrence of sizeable condensate fraction and density modulation can be interpreted as the analogous, in a quasicrystalline lattice, of supersolid phases occurring in conventional periodic lattices. For sufficiently large interaction strength and particle density, global condensation is lost and quantum exchanges are restricted to specific spatial regions. The emerging quantum phase is therefore a Bose Glass, which here is stabilized in the absence of any source of disorder or quasidisorder, purely as a result of the interplay between quantum effects,  particle interactions and quasicrystalline substrate. This finding clearly indicates that (quasi)disorder is not essential to observe Bose Glass physics. Our results are of interest for ongoing experiments on (quasi)disorder-free quasicrystalline lattices.
\end{abstract}

\maketitle

Quasicrystals are structures whose constituents are arranged in ordered but not periodic patterns \cite{jan12, pen74}. While originally discovered in solid-state materials \cite{PhysRevLett.53.1951, Bindi2009}, their peculiar geometrical properties \cite{lev86, sen96} have inspired extensive investigation in statistical and quantum many-body physics. In this framework, great interest has been elicited by many-body systems on quasicrystalline substrates (i.e., lattices), which have been experimentally studied using photonic platforms \cite{Vardeny:2013aa, RevModPhys.93.045001, Bandres2016, 2211.06047}, relativistic Dirac fermions \cite{Ahn2018}, and electronic states of layered graphene \cite{PhysRevB.99.165430}, to cite a few examples.
Bosonic and fermionic systems such as atoms in trapping potentials and/or cavities have also recently emerged as a highly controllable and versatile setup for the realization of quasicrystalline physics \cite{PhysRevA.72.053607, Schreiber2015, gau21, Sbroscia2020,Viebahn2019, PhysRevA.105.L011301, PhysRevB.105.134521, PhysRevB.101.134522, PhysRevLett.119.215304, PhysRevLett.123.210604, 2210.05691, PhysRevLett.120.060407, PhysRevLett.111.185304}. In these experiments, the superposition of incommensurate laser standing waves is used to engineer effective quasicrystalline lattices. In particular, the quasicristalline arrays generated via this procedure are characterized by site-dependent depths of the potential minima defining the lattice sites. This peculiar feature, referred to as \textit{quasidisorder}, is deterministic, and hence different from standard disorder, which is intrinsically related to some degree of randomness (e.g., between different system realizations) \cite{PhysRevLett.91.080403,PhysRevLett.98.130404,Schreiber2015}.

From the theoretical point of view, the investigation of Hamiltonians on quasicrystalline substrates pointed out possible intriguing connections between quasicrystalline physics and elasticity theory \cite{PhysRevX.11.041051}, topological properties \cite{Fan2021, PhysRevLett.124.036803,Kraus2016, PhysRevLett.123.196401, PhysRevLett.125.265702,PhysRevB.100.214109}, and superconductivity \cite{PhysRevLett.116.257002, PhysRevB.104.144511}. Clearly, the accurate determination of the low-temperature ($T$) phases hosted by these systems, as well as of the mechanisms underlying the stabilization of such phases, constitutes a point of primary importance.

\begin{figure}[t!]
  \begin{center}
    \includegraphics[width=\linewidth]{%
      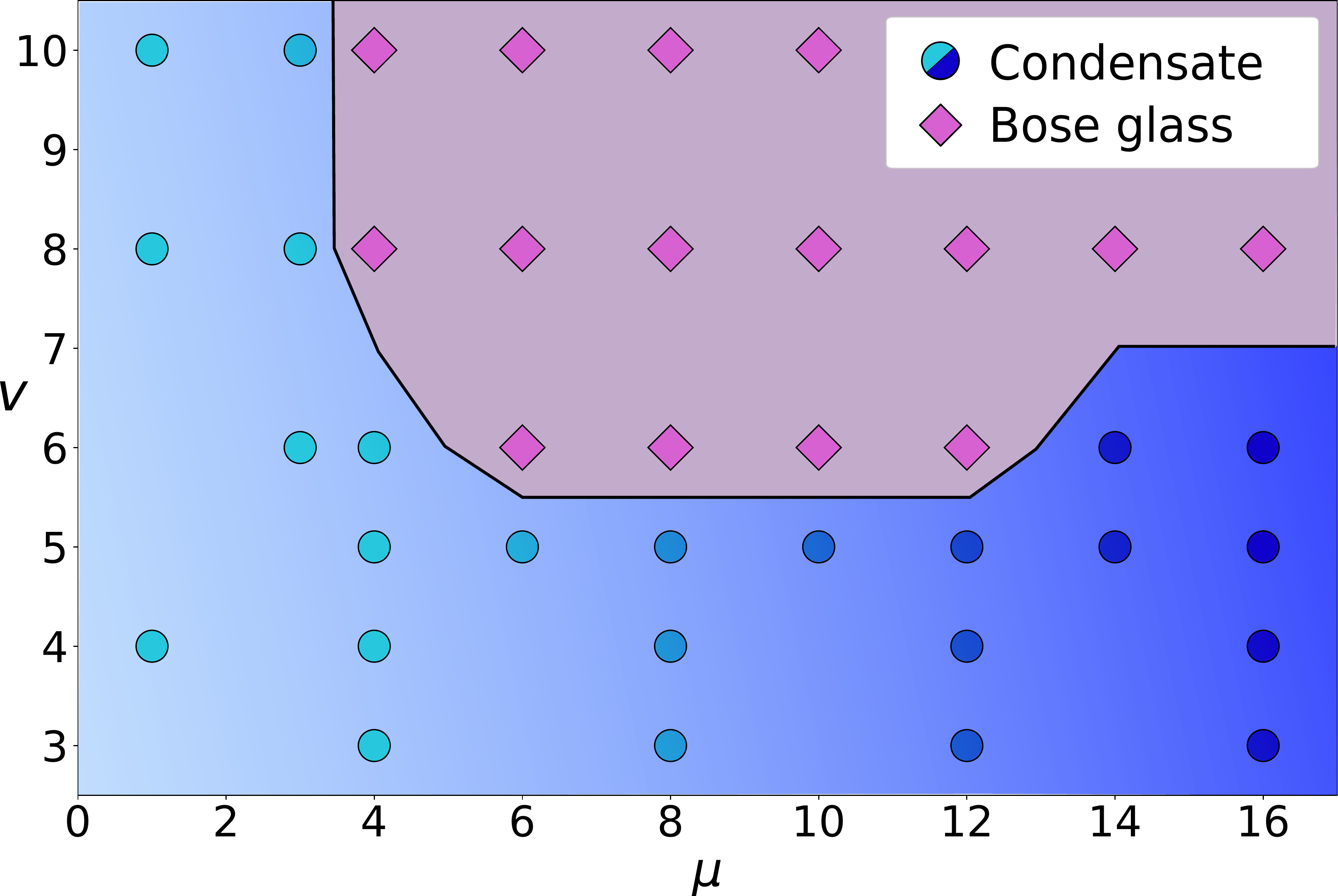}
    \caption{%
      Low-temperature phase diagram of the model in eq.~\eqref{eq:hamiltonian} as a function of the interaction strength $V$ and of the chemical potential $\mu$. Performed simulations converging to condensate (Bose glass) states are marked by circles (diamonds). The gradient from light to dark color indicates increasing density modulation strength in the condensate.
    }
    \label{fig1}
  \end{center}
\end{figure}

Previous theoretical studies \cite{gau21,PhysRevA.105.L011301,e24020265,joh21}, inspired by current cold atom setups, focused on (quasi)disordered models hosting at low $T$ superfluid and/or condensate, insulating, and Bose glass (BG) states. The latter, usually arising in disordered scenarios \cite{Fisher1989, Cepr2007, Cepr2011, Cepr2010, soy11, Yu2012a, Cepr2014, Cep2016}, are exotic insulators which feature rare regions of delocalized particles involved in local exchange cycles. Whether or not the typical low-$T$ phase diagram of models on quasicrystalline lattices changes if (quasi)disorder is lifted is a crucial, still open question. Providing an accurate answer to this question is i) fundamental to unveil the conjectured relation (if any) between the intriguing BG phase and (quasi)disorder, and ii) of direct relevance for possible realizations of disorder and quasidisorder-free quasicrystalline Hamiltonians in, e.g., photonic systems \cite{Bandres2016}.

In this work, we investigate the low-$T$ physics of a (quasi)disorder-free, extended hardcore-boson Hubbard model on a two-dimensional quasicrystalline lattice. By means of numerically exact quantum Monte Carlo simulations, we determine the phase diagram of our system of interest as a function of both the chemical potential and the interaction strength (see Fig.~\ref{fig1}). We find that for weak interactions and/or low particle density (i.e., low chemical potential) the system is a homogeneous Bose-Einstein condensate; as particle number increases, this state develops density modulations, similarly to what happens in a superfluid-supersolid transition for models on Bravais lattices \cite{RevModPhys.84.759}. Conversely, for sufficiently large interactions, the condensate makes way for a BG phase, where patches of exchanging particles survive in a globally insulating, yet compressible state. The demonstration that a BG can be stabilized in the absence of (quasi)disorder, uniquely as a result of  the interaction between particles and the quasicrystalline nature of the system substrate, is the central result of our work.

The Hamiltonian considered in the present study reads
\begin{equation}
  H = - \sum_{\langle i,j \rangle} \left( b^{\dagger}_i b_j +
  \mathrm{h.c.} \right) - \mu \sum_{i = 1}^N n_i + V
  \sum_{\langle i,j \rangle} n_i n_j,
  \label{eq:hamiltonian}
\end{equation}
where $b_i$ and $b_i^{\dagger}$ are the annihilation and
creation operator for a hardcore boson on site $i$,
respectively, $n_i \equiv b^{\dagger}_i b_i$, $\mu$ is the
chemical potential, $N$ is the number of sites, $V > 0$ is the
interaction strength, and $\langle i,j \rangle$, as detailed below, denotes all
couples of sites connected by the hopping and interaction
terms [i.e., the first and the last term in eq~\eqref{eq:hamiltonian}, respectively]. In this Letter, the coefficient of the hopping term is set to 1, and chosen as unit of energy and temperature.
\begin{figure}[t!]
  \begin{center}
    \includegraphics[width=\linewidth]{%
      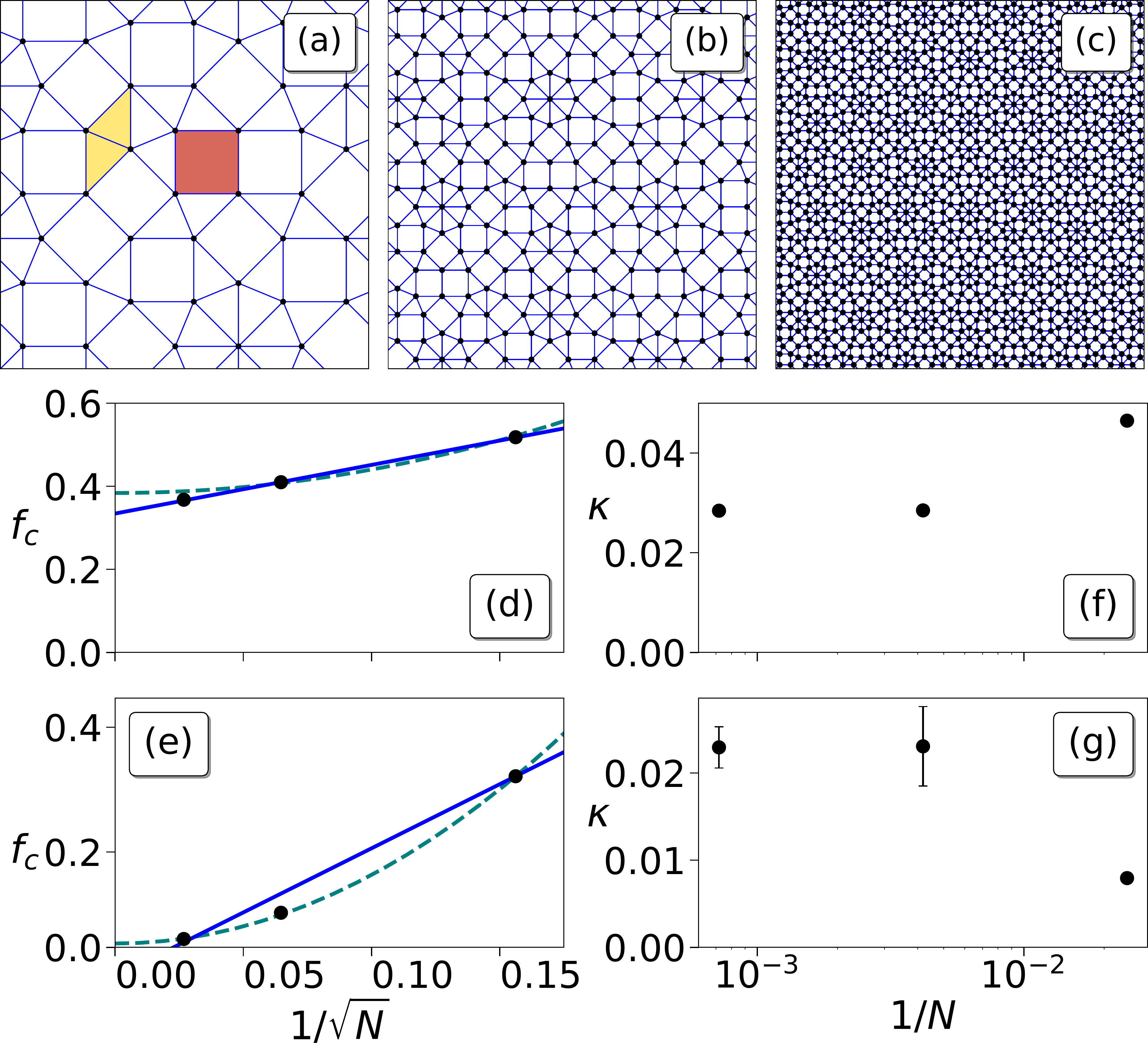}
    \caption{%
      (a--c): Approximants of the Ammann-Beenker tiling for $N = 41$ (a), $N = 239$ (b), and $N = 1393$ sites (c). Dots denote the lattice sites, and segments are drawn between pairs of sites connected by the two-body Hamiltonian terms. The red and yellow areas are a square and rhombic tile in the Ammann-Beenker tiling, respectively. (d--e): lattice-size dependence of the condensate fraction $f_c$ for $\mu = 1, V = 4$ (d) and $\mu = 4, V = 8$ (e). Error bars are smaller than the symbol size. Solid and dashed lines correspond to linear and quadratic (in $N^{-1/2}$) fitting functions to our numerical data, respectively. Our estimated value of $f_c$ in the thermodynamic limit is, for each given $\mu$ and $V$, the one resulting from the fit yielding the lowest reduced chi-square. (f--g): lattice-size dependence of the compressibility  $\kappa$ for the same parameters of panel (d) and (e), respectively; $\kappa$ stays finite in the thermodynamic limit (convergence within error bars is achieved for the two larger sizes).}
    \label{fig2}
  \end{center}
\end{figure}
Our adopted quasicrystalline substrate is realized by taking as lattice sites the vertices of approximants of the Ammann-Beenker tiling \cite{gru13, dun89}. These structures, obtained here via the inflation mapping method \cite{dun89}, are only defined for specific sizes: in this work, we analyze the three consecutive sizes $N = 41, 239, 1393$ [the corresponding lattices are shown in Fig.~\ref{fig2}(a--c)].
We consider as connected by the two-body terms of the Hamiltonian all pairs of sites at a distance equal or smaller than the side of the square tiles in the approximants [filled square in Fig.~\ref{fig2}(a)].
The segments in Fig.~\ref{fig2} explicitly illustrate the connectivity of each site \cite{nota1}.

The interparticle interaction in eq.~\eqref{eq:hamiltonian} is of relevance for experiments with cold atoms, for example, in Rydberg-dressing schemes where it has already been realized with fermions \cite{PhysRevX.11.021036} while, in the bosonic case, it has been the object of different theoretical/numerical studies leading, for various choices of Bravais lattice and interaction range,  to the prediction of exotic states of matter in both the equilibrium and out-of-equilibrium regime \cite{Batrouni2000, Wessel2005, Boninsegni2005, AnMePu2016, MaMePu2019, Angelone2020}.
Notably, our model is also of interest for current experiments with Rydberg atoms in the resonant regime, which may realize essentially any lattice geometry by means of optical tweezers, and a spin-$1/2$ XXZ Hamiltonian as that in eq.~\eqref{eq:hamiltonian}, albeit with a ``tail" algebraically decaying as $r^{-\gamma}$, with $\gamma=3$ or $6$ \cite{PhysRevX.11.011011, PRXQuantum.3.020303}.

We study eq.~\eqref{eq:hamiltonian} via Path Integral Monte
Carlo (PIMC) simulations based on the worm algorithm for
lattice systems in the grandcanonical ensemble \cite{Prokofev:1998aa}. In this scheme, each quantum particle is mapped into a classical polymer, known as \textit{worldline}. Hence, the original, $d$-dimensional quantum system is exactly transformed into a $(d+1)$-dimensional classical counterpart. Statistical mean values of operators (denoted by $\ave{\cdots}$ in the following) are then computed by averaging over both the additional synthetic dimension (i.e., \textit{imaginary time}) and the stochastically sampled classical configurations.

We determine the phase diagram of the Hamiltonian in eq.~\eqref{eq:hamiltonian} at temperatures $T = 1/4$ and $1/8$, observing no relevant discrepancies between the obtained results in the two cases. All figures shown consistently refer to the choice $T = 1/8$. Figure~\ref{fig1} displays our computed phase diagram, which hosts both condensate and insulating phases occurring at low and high (for sufficiently large interaction strength) values of chemical potential, respectively. Interestingly, at $V=6$, we find that the condensate phase extends up to $\mu \simeq 5$, ``re-entering" for $\mu \gtrsim 14$. The order parameter used to discriminate between condensate and insulating phases is the condensate fraction $f_c$. The latter is defined as the maximum among the relative occupations of the single-particle Hamiltonian eigenstates: hence, it will be finite (vanishing) in condensate (insulating) states in the thermodynamic limit. Operatively, $f_c$ can be estimated as the largest eigenvalue of the renormalized one-body density matrix $G_{ij} \equiv \ave{b_i^{\dagger} b_j}/N$ \cite{Penrose1956}. Our finite-size results are then extrapolated in the $N \to \infty$ limit to estimate the condensate-insulator phase boundary (solid line in Fig.~\ref{fig1}). 
Figure~\ref{fig2} shows examples of this procedure where $f_c$ scales to finite [panel (d)] or vanishing [panel (e)] values; the corresponding points in the phase diagram are then classified as condensate and insulating, respectively.

We also estimate the compressibility $\kappa = N\beta (\langle n^2 \rangle - \langle n \rangle^2)$, where $n$ is the average particle occupation per site of our simulated system. A finite value of $\kappa$ occurs both in a condensate and in a (globally insulating) BG where condensation occurs locally in rare regions (see below). We find that $\kappa$ stays  finite in the thermodynamic limit in the whole hamiltonian-parameter range explored in this work. The size dependence of $\kappa$ for the same parameter sets of panels (d) and (e) is shown in panels (f) and (g), respectively, The latter illustrate in particular that  across our condensate-to-insulator phase transition the compressibility does not vanish. The value of  $\kappa$, as expected, decreases deep in the insulating phase remaining however finite. The insulating phase of our phase diagram is therefore a BG.

\begin{figure}[t!]
  \begin{center}
    \includegraphics[width=\linewidth]{%
      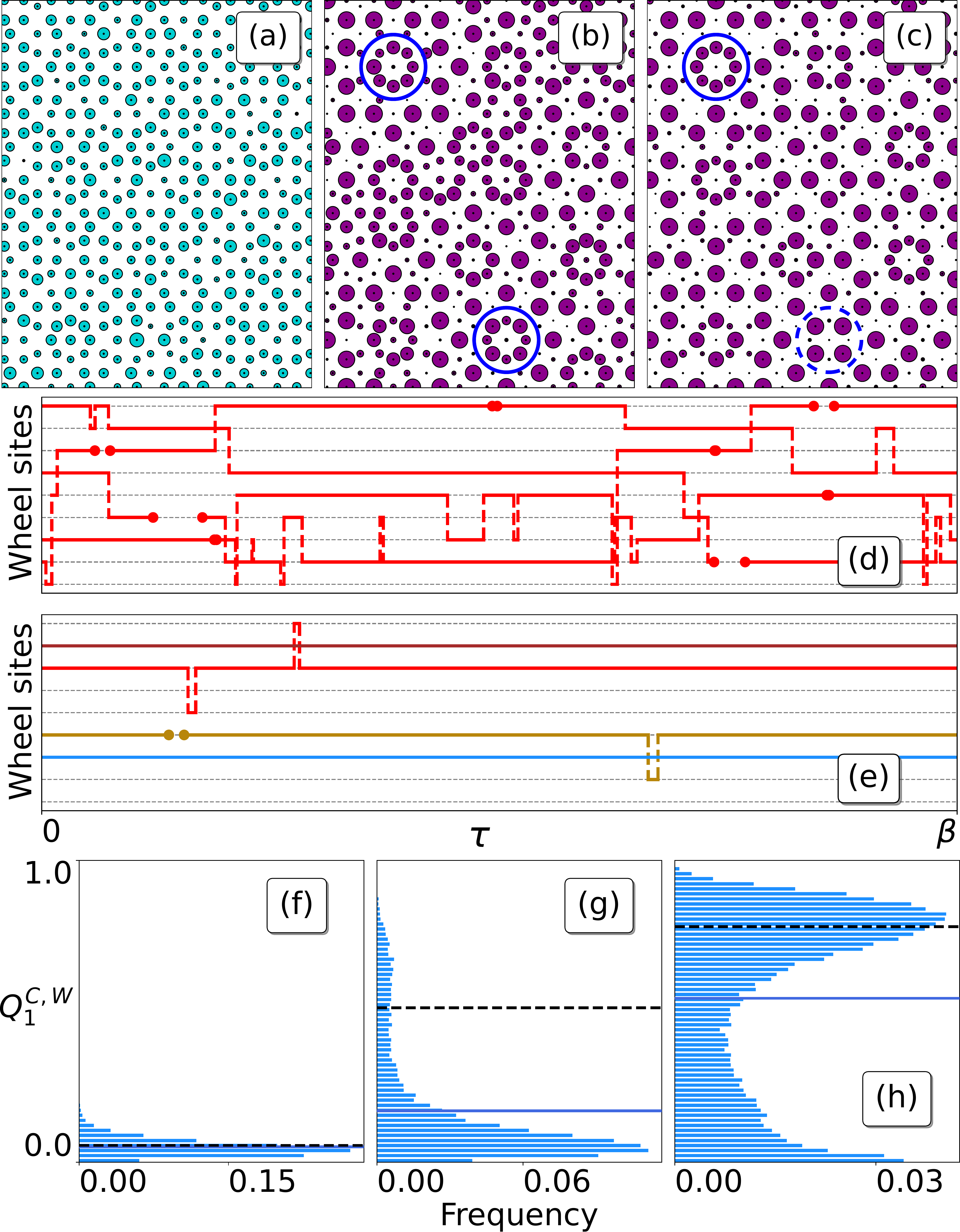}
    \caption{%
      (a--c): Imaginary-time averaged density maps for a given PIMC configuration. Panel (a) refers to a condensate ($\mu = 4, V = 5$), while panels (b) and (c) refer to Bose glass states ($\mu = 8, V = 6$ and $\mu = 8, V = 8$, respectively). Each small black dot represents a lattice site, while the size of the corresponding colored circle is proportional to its occupation. Solid and dashed circles point out examples of quantum and classical wheels, respectively (see text). (d--e): Worldline structure of the particles occupying a quantum (d) and classical (e) wheel. Horizontal dashed lines denote the sites in the wheel, vertical ones correspond to particle hoppings, and small circles denote hopping outside of the wheel. Separate worldlines are shown in different colors. (f--h): Histograms for the collected values of $Q_1^{C,W}$ (see text) in all wheels and configurations. Panel (f--h) are obtained for the same parameter values of panels (a--c), respectively. The horizontal solid (dashed) line indicates the value of $Q_1$ ($Q_1^{\mathrm{wheels}}$) (see text).
    }
    \label{fig3}
  \end{center}
\end{figure} 

Further insight into the obtained phases can be gained by examining single-configuration density maps, such as those displayed in Fig.~\ref{fig3}(a--c). Here, lattice sites are marked by black dots, while the sizes of the corresponding colored circles are proportional to the site occupation averaged over imaginary time in a single PIMC configuration. In a condensate [panel (a)], essentially all sites have a non-integer occupation, signaling particle delocalization. This has to be contrasted with a classical configuration, in which particles do not change their position in imaginary time, yielding only $0$ or $1$ average occupations.

Increasing the interaction strength and/or the chemical potential hampers delocalization, leading from condensate to insulating states. The nature of such insulating states is however nontrivial, as one can realize via inspection of single-configuration density maps [Fig.~\ref{fig3}(b--c)]. Here, intriguing physics occurs in ``wheels'', i.e., eightfold rotationally-symmetric groups of lattice sites. Wheel centers are the only sites in the quasicrystal with exactly eight nearest neighbors, which allows us to easily identify them. Remarkably, these structures may host delocalized particles, in contrast with the substantially classical behavior of the rest of the system: all this is again apparent from the noninteger site occupations of these ``quantum wheels'' [solid circles in Fig.~\ref{fig3}(b--c)], as opposed to the essentially integer ones  (i.e., 0 or 1) observed elsewhere. ``Classical wheels", where particles are not delocalized in the quantum sense, may also coexist with their quantum counterparts [dashed circle in Fig.~\ref{fig3}(c)].

With the aim of investigating the presence of particle exchanges in quantum wheels, we inspect the worldline configurations of particles residing in these regions. Indeed, in the path integral representation, worldlines of particles involved in exchange cycles fuse into one, consistently with the notion of indistinguishability. Figure~\ref{fig3}(d) displays an example of worldline structure in a quantum wheel. Here, all particles are involved in a single exchange cycle, which covers the entire wheel but essentially does not extend outside of it. Conversely, the worldlines in a classical wheel [Fig.~\ref{fig3}(e)] are basically flat, as expected for classically-behaving particles. In quantum wheels particles are then delocalized and involved in quantum exchanges; however, due to the local (i.e., confined to the wheel region) character of these permutation cycles, global condensation, which is associated to system-wide exchange cycles, is suppressed. The emerging physical picture is therefore that of a globally insulating compressible state hosting condensate patches, namely, a BG.

In order to quantify the presence of quantum wheels in our observed insulating
states, we introduce the single-configuration density modulation parameter $Q_1^C$, defined for a generic PIMC configuration $C$ as
\begin{equation}
  Q_1^C \equiv \frac{\sum_i \left( n_i^C - \ol{n_i^C}
  \right)^2}{N \ol{n_i^C} \left( 1 - \ol{n_i^C} \right)} \equiv q \left( \left \{ n_i^C \right \} \right).
  \label{eq:q1c}
\end{equation}
Here, $\left \{ n_i^C \right \}$ is the set of the imaginary-time-averaged occupations in $C$, and $\ol{n_i^C} = N^{-1} \sum_i n_i^C$. The $Q_1^C$ parameter is inspired by similar ones employed in the context of glassy physics \cite{Edwards1975, Carleo2009, AnMePu2016, Angelone2020}, and here measures the inhomogeneity of the imaginary-time-averaged site occupations in a single configuration. Specifically, its extremal values $0$ and $1$ correspond to perfectly delocalized configurations (i.e., uniform $n_i^C < 1$) and perfectly classical particles (i.e., each of the $n_i^C$ is either $0$ or $1$), respectively. $Q_1^C$ can be computed either for the entire system or for a subset of lattice sites, by appropriately restricting the sums. In the following, we will use the notation $Q_1^{C,W}$ to indicate the value of this observable computed on a wheel $W$ in a single configuration $C$.

In Fig.~\ref{fig3}(f--h) we show histograms, accumulated over the entire PIMC simulation, of the values of $Q_1^{C,W}$ for all wheels. In both condensate [panel (f)] and insulating states [panels (g--h)], the histograms display a peak close to $0$, which denotes the presence of a large number of quantum wheels. In turn, classical wheels are associated to high values of $Q_1^{C,W}$. The latter are rather infrequent in BG states close to the condensate-BG transition (see, e.g., the histogram in panel (g)). Conversely, at higher values of $V$ and/or $\mu$ [panel (h)] classical wheels become more frequent, and the histogram develops a second peak at high values of $Q_1^{C,W}$. However, the low-$Q_1^{C,W}$ peak remains, signaling the persistence of quantum wheels.

 The nature of the wheels can be compared to that of the rest of the system by computing the configuration average of $Q_1^C$ for the whole system, which we dub $Q_1$ (dashed lines in Fig.~\ref{fig3}) and the average of $Q_1^{C,W}$ over all configurations and wheels, which we dub $Q_1^{\mathrm{wheels}}$ (solid lines in the same figure). We find that these two averages are essentially coinciding in the case of a condensate [panel (f)], while in our insulating states they display a sizeable difference [panel (g--h)]. This implies that in our globally non-condensate states a significant fraction of the wheels host delocalized particles.
For all parameter sets associated to insulating states in our phase diagram, the obtained values of the relative difference $W \equiv (Q_1 - Q_1^{\mathrm{wheels}})/Q_1$ stays finite in the $N \to \infty$ limit, indicating the presence in the system of a sizeable number of quantum wheels. The latter are responsible, in our BG phase, for the finite value of the compressibility.

In previous studies (see for example Refs. \cite{gau21, Sbroscia2020}) the condensate-BG transition has been investigated in continuous-space models  explicitly including quasidisorder. There, a BG phase emerges in the limit of low particle density, weak interparticle potential and sufficiently large strength of quasidisorder, while, away from the weakly interacting limit, i.e., for increasing interparticle potential, the BG progressively makes way to a Mott insulator state. In the present study the situation is substantially different: in the dilute, weakly interacting regime, our system is a homogeneous condensate and the transition to a BG is not driven by quasidisorder, which is absent in our model. Remarkably, our observed BG phase occurs for sufficiently high particle density, in a regime where the interparticle potential takes values $V \sim 6$--$10$. Such a BG is ultimately stabilized by a subtle interplay between interactions, quantum effects, and  quasicrystalline lattice.

\begin{figure}[t]
  \begin{center}
    \includegraphics[width=\linewidth]{%
      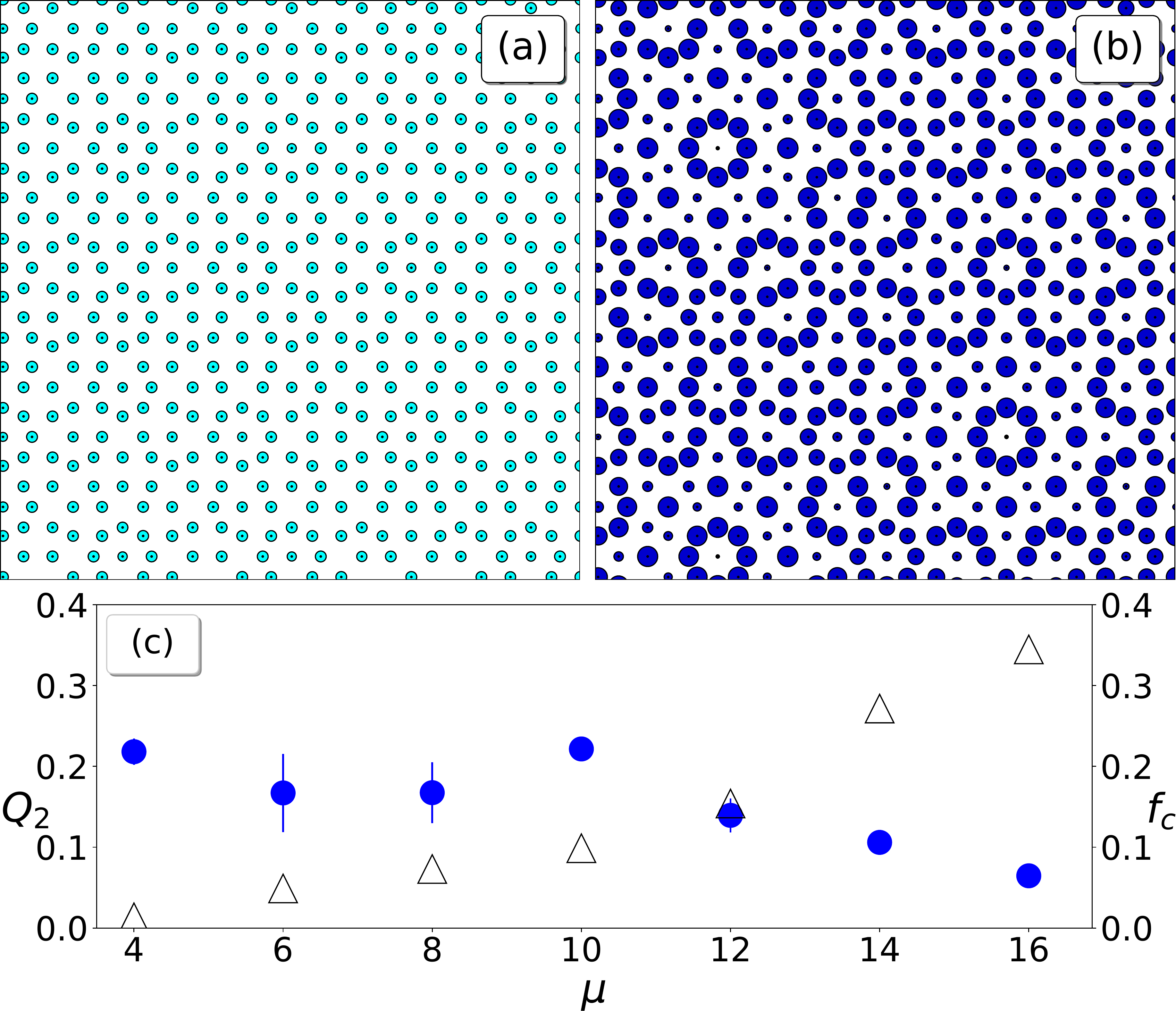}
    
    \caption{%
      (a--b): Statistically averaged density maps for a portion of a $N = 1393$ lattice, at $V = 5, \mu = 4$ (a) and $V = 5, \mu = 16$ (b). Each small black dot represents a lattice site, while the size of the corresponding colored circle represents its occupation. 
      (c): Values of $f_c$ (circles) and $Q_2$ (triangles) extrapolated to the thermodynamic limit (see text). Error bars, where not visible, are smaller than the symbol size.
    }
    
    \label{fig4}
  \end{center}
\end{figure} 

The analysis of PIMC density maps also allows us to characterize nontrivial features in our obtained condensate states. Specifically, in Fig.~\ref{fig4}(a--b) we show stochastically averaged density maps for two condensate states, at weak interactions and low and high particle density, respectively. In the former case the density map, as expected, is basically uniform [panel (a)]; conversely, in the latter, sizeable density inhomogeneities are evident [panel (b)]. 

We measure the strength of these density modulations through the parameter defined in eq.~\eqref{eq:q1c} computed on the averaged density map, $Q_2 \equiv q \left( \left \{ \ave{n_i} \right \} \right)$. This parameter is expected to be $0$ in a perfect condensate, where statistically-averaged site occupations are uniform, while it is finite in the presence of density modulations.

Figure~\ref{fig4}(c) shows values of $Q_2$ (triangles) at interaction strength $V=5$ and increasing $\mu$ (i.e., increasing particle density). We see how this observable grows monotonically with $\mu$, reaching up to $\sim 30\%$ for $\mu = 16$. Our $Q_2$ estimates remain substantially unchanged when $N$ increases from $239$ to $1393$, i.e., the largest size considered in our study. Therefore, the observed density inhomogeneities will persist in the thermodynamic limit. Concomitantly with the development of density fluctuations, the value of the condensate fraction extrapolated to the thermodynamic limit (circles) decreases from approximately 0.2 to slightly less then 0.1, hence $f_c$ stays finite in the whole
parameter range of Fig.~\ref{fig4}(c). The coexistence in our system of global condensation and density modulations is reminiscent of supersolid states on Bravais lattices \cite{RevModPhys.84.759}.

In conclusion, we have studied the low-temperature physics of a model of hardcore bosons on a two dimensional quasicrystalline lattice, demonstrating that a Bose glass state may be stabilized in the absence of either disorder or quasidisorder sources. Indeed, we show that the latter are not essential ingredients for the appearance of a Bose glass, which may result purely from the interplay between quantum effects, interactions and the non-periodic nature of a quasicrystalline lattice. 
We also find that the homogeneous condensate phase characterizing our Hamiltonian at low interaction strength and chemical potential (i.e., particle density) progressively develops sizeable density modulation for increasing values of $\mu$. The resulting modulated condensate calls for an intriguing analogy with supersolid states predicted in a variety of systems on periodic lattices. 

Our results contribute to motivate an experimental focus towards the development of quasidisorder-free quasicrystalline lattices, for example in photonic systems \cite{Vardeny:2013aa}, or in different kinds of ultracold-atom frameworks \cite{GAUTHIER20211, PhysRevX.11.011011, PRXQuantum.3.020303}. 
An interesting extension of our work is its generalization to different types of interactions, such as dipolar ones, which have been associated in recent years to a wealth of exotic physical phenomena \cite{Baranov2008, Chomaz2022, PRXQuantum.3.020303}. 
A detailed analysis of bosonic physics across a variety of quasicrystalline geometries--and even correlated disorder--also constitutes a viable option for further investigations e.g., in the framework of hyperuniform lattices \cite{torquato2003local, TORQUATO20181}.

\begin{acknowledgments}
We thank the NICIS-CHPC agency for providing computational resources. F. M. is supported by ANR (``EELS"
project) and QuantERA (``MAQS" project). M. C. and F. C. acknowledge financial support from PNRR MUR Project
No. PE0000023-NQSTI.
\end{acknowledgments}

%\bibliography{bose.bib}

\begin{thebibliography}{68}
\bibitem [{not()}]{nota1}%
  \BibitemOpen
  \href@noop {} {}\bibinfo {howpublished} {The number of neighbors each site is
  connected to by our Hamiltonian terms is not identical for all sites.
  However, we remark that this effect is fundamentally different from
  (quasi)disorder, which would be implemented by, e.g., site-dependent chemical
  potentials. Indeed, our connectivity variance is an unavoidable consequence
  of the choice of a quasicrystalline geometry.}\BibitemShut {Stop}%
\end{thebibliography}

%

\end{document}